\documentclass[prl,twocolumn,amsmath,amssymb,floatfix]{revtex4-1}
\pdfoutput=1
\usepackage{mathtools}
\usepackage{tensor, color}
\usepackage{braket}
\usepackage{graphicx}
\usepackage{bbold}
\usepackage{bm}
\DeclareMathOperator{\tr}{tr}

\newcommand{\be}{\begin{equation}}
\newcommand{\ee}{\end{equation}}
\newcommand{\beq}{\begin{eqnarray}}
\newcommand{\eeq}{\end{eqnarray}}
\newcommand{\bml}{\begin{multline}}
\newcommand{\eml}{\end{multline}}

\begin{document}

\title{$SU(3)$ semiclassical representation of quantum dynamics of interacting spins} \author{Shainen M Davidson} \email[Corresponding author: ] {shainen@bu.edu} \author{Anatoli Polkovnikov} \affiliation{Department of Physics, Boston University, Boston, MA 02215, USA} \date{\today}

\begin{abstract}

  We present a formalism for simulating quantum dynamics of lattice spin-one systems by first introducing local hidden variables and then doing semiclassical (truncated Wigner) approximation in the extended phase space. In this way we exactly take into account the local on-site Hamiltonian and approximately treat spin-spin interactions.  In particular, we represent each spin with eight classical $SU(3)$ variables. Three of them represent the usual spin components and five others are hidden
  variables representing local spin-spin correlations. We compare our formalism with exact quantum dynamics of fully connected spin systems and find very good agreement. As an application we discuss quench dynamics of a Bose-Hubbard model near the superfluid-insulator transition for a 3D lattice system consisting of 1000 sites.

\end{abstract}

\maketitle

Recent experiments in such areas as ultra-cold gases, coupled atom-photon systems, ultrafast pump-probe spectroscopy in solids, and others have stimulated active theoretical research in quantum dynamics of interacting systems. While there has been significant progress in various directions, our understanding is still quite limited. Partly this is due to a lack of reliable and controllable numerical methods. Perhaps the most powerful methods in equilibrium based on quantum Monte-Carlo are not
readily adopted to non-equilibrium systems. The other available methods include: simulations based on exact diagonalization; dynamical renormalization group and related matrix product states methods~\cite{Scholwock2005}; dynamical mean field theory based methods~\cite{Aoki2013, Trefzger2011} mostly developed for fermions and only recently applied to bosons~\cite{Strand2014}; quantum kinetic equations and Keldysh diagrammatic technique~\cite{Kamenev2009}; and phase space methods. The latter has
recently become a major tool for studying dynamics of various systems, from interacting atomic clocks to the early Universe~\cite{Steel1998, Blakie2008, Altland2009, Lancaster2010, Polkovnikov2010a, Opanchuk2013, Rey2014}. These methods are based on mapping the density matrix and operators into classical functions, which depend on phase space variables like coordinates and momenta, complex wave amplitudes, or classical spin degrees of freedom.

Phase space methods are very efficient for systems near the classical or non-interacting limit, where the quantum evolution is well described by classical trajectories. The main idea of the present work is that we can significantly improve the accuracy of these methods by extending classical phase space by introducing new (hidden) phase space variables. In this way we can take into account local quantum fluctuations exactly and treat other degrees of freedom approximately. We illustrate this
idea by focusing on an example of spin-one coupled systems where, in addition to three variables representing the $x, y, z$ components of the spin, we introduce five additional hidden variables representing local spin-spin correlations. We believe these ideas can be further extended and applied to study a large class of interacting systems both in and out of equilibrium.

Before proceeding with our ideas we briefly review the phase space representation of quantum dynamics of interacting bosonic systems. For concreteness we focus on the Wigner-Weyl quantization (see Supplementary Material for a brief overview). Any operator of a quantum system that we would normally represent through a function of boson operators $\hat a$ and $\hat a^\dagger$ can be mapped to a function over the classical phase space of (complex) canonical variables $\alpha$ and
$\alpha^*$. This function is called the Weyl symbol of the operator and it is uniquely defined. The Weyl symbol of the density matrix is known as the Wigner function~\cite{Hillery1984, Polkovnikov2010a}. It plays the role of the (quasi)-probability distribution. In general, time evolution of the Wigner function is given by a Fokker-Planck equation with high derivatives, which is hard to handle. However, near the classical limit or for non-interacting systems one can use the so-called truncated
Wigner approximation (TWA)~\cite{Steel1998, Blakie2008, Polkovnikov2010a}, where the Wigner function, like a classical probability distribution, is conserved on classical trajectories. Then the expectation value of any operator can be straightforwardly computed at any moment in time: \be \langle \hat \Omega(t)\rangle\approx\int d\vec\alpha_0 d\vec\alpha_0^\ast W(\vec\alpha_0,\vec\alpha_0^\ast) \Omega_W(\vec\alpha_{cl}(t),\vec\alpha ^\ast_{cl}(t)), \ee where $W(\vec\alpha_0,\vec\alpha_0^\ast)$ is
the Wigner function representing the initial state of the system, $\Omega_W(\vec\alpha,\vec\alpha^\ast)$ is the Weyl symbol of the operator $\hat \Omega$, and the classical paths are found by solving Hamilton's equations,
\begin{align}
  \label{classeq}
  i \dot \alpha_{cl}=\frac{\partial H_W}{\partial \alpha_{cl}^\ast}.
\end{align}
TWA becomes more accurate as the number of particles increases. It is always exact when the Hamiltonian is non-interacting, i.e. quadratic in boson operators.

One way to implement TWA with spin operators is to use the Schwinger boson representation for spin, where the spin operators are built from 2 creation/annihilation pairs of boson operators contracted with the Pauli matrices:
\begin{align}
  \label{eq:1}
  \hat S_\alpha = \hat a^\dagger_i \frac{1}{2} \sigma_\alpha^{ij} \hat a_j.
\end{align}
The boson pairs ``inherit'' the commutation relations of the matrices they are contracted with, in this case the spin algebra of $SU(2)$. Representing the spin Hamiltonian in this way, we can use the machinery of coherent state TWA to approximate the dynamics (see~\cite{Polkovnikov2010a} Section 3.3). Since TWA dynamics is exact for a Hamiltonian that is quadratic in boson operators, it is also exact for a system linear in spin operators.

We can similarly represent any $SU(N)$ group of operators with commutation relations $[\hat X_\alpha, \hat X_\beta]=i f_{\alpha\beta\gamma} \hat X_\gamma$ using $N$ boson operator pairs and the $N$ dimensional matrices $T^{ij}_\alpha$ with the corresponding algebra:
\begin{align}
  \label{genrep}
  \hat X_\alpha=\hat a^\dagger_i T_\alpha^{ij} \hat a_j.
\end{align}
Note that any Hamiltonian that acts on a Hilbert space of dimension $N$ (or indeed any Hermitian operator of dimension $N$) can be represented as a linear superposition of the generators of the $SU(N)$ group and an identity matrix. Again, the TWA dynamics of any Hamiltonian linear in $SU(N)$ variables will be exact. This statement can be alternatively understood based on the linearity of the Heisenberg equations of motion as was recently explored in Ref.~\cite{Galitski2011}.

Taking the Wigner-Weyl transform of \eqref{genrep}, we can define real number valued variables
\begin{align}
  \label{classgenrep}
  X_\alpha=\alpha^*_{i} T_\alpha^{ij} \alpha_{j},
\end{align}
and the classical equations of motion for the canonical variables \eqref{classeq} will lead to dynamics for these variables as (see Supplementary Material for details)\be \dot X_\alpha^{cl}=f_{\alpha\beta\gamma} {\partial H_W\over \partial X^{cl}_\beta} X^{cl}_\gamma.
\label{bloch_group}
\ee

In the case of the $SU(2)$ group representing e.g. spin in a magnetic field the structure constants are given by $f_{\alpha\beta\gamma}=\epsilon_{\alpha\beta\gamma}$, where $\epsilon$ is the fully antisymmetric tensor. Then it is easy to see that the equation above represents the standard Bloch equations $\dot{\vec X}=\vec X\times \vec B$, where $H=-\vec B\vec X$ and $\vec B$ is a (possibly time-dependent) magnetic field.

If we have a single spin 1/2 degree of freedom then its Hamiltonian can always be represented as a linear superposition of Pauli matrices (generators of the $SU(2)$ group) and thus semiclassical dynamics are exact. Consider now the slightly more complicated situation of an isolated spin-one degree of freedom. A generic Hamiltonian for spin-one can include interactions, i.e. terms non-linear in spin operators. Just to be specific consider the interactions of the type $\hat S_z^2$ such that \be
\hat H_{I}=-\vec B\vec {\hat S}+(U/ 2) \hat S_z^2.
\label{one_spin_hamiltonian}
\ee Note that both $\vec B$ and $U$ can explicitly depend on time. If the interaction coupling $U$ is zero we are back to the situation discussed before, where the Hamiltonian is a linear superposition of $SU(2)$ generators and semiclassical dynamics are exact. However, for finite interactions the semiclassical approximation breaks down at long times. However, since spin-one operators exist in a 3 dimensional Hilbert space, we can make the semiclassical dynamics exact by enlarging the group to
$SU(3)$ by using the 8 3-dimensional matrix generators of $SU(3)$ as the matrices $T_\alpha$ in \eqref{genrep}. Those can be chosen to be Gell-Mann matrices, but it is more convenient to use linear combination such that the first three generators are the spin-one spin matrices~\cite{Kiselev2013}, e.g.
\begin{align}
  \label{eq:3}
  T_1 =
  \begin{pmatrix}
    0 & \frac{1}{\sqrt{2}} & 0 \\
    \frac{1}{\sqrt{2}} & 0 & \frac{1}{\sqrt{2}} \\
    0 & \frac{1}{\sqrt{2}} & 0
  \end{pmatrix}
\end{align}
and thus $\hat X_1=\left(\hat a_1^\dagger \hat a_2 +\hat a_3^\dagger \hat a_2 +h.c.\right)/\sqrt{2}$ and so on (see Supplementary Material for details).

Because any Hermitian $3\times 3$ matrix can be expressed through a linear combination of $SU(3)$ matrices (and the identity) the interaction term also becomes linear in the $SU(3)$ representation. For example, for our choice of $SU(3)$ generators we have $\hat S_z^2 = (2 - \sqrt{3} \hat X_8)/3$.

Thus the whole spin-1 Hamiltonian becomes linear in the generators of $SU(3)$ and the equations of motion~(\ref{bloch_group}) become exact. These equations can be interpreted as an exact classical representation of quantum dynamics of a spin-one system in the eight-dimensional phase space spanned by the classical phase space variables $X_1,\dots, X_8$\footnote{That the dynamics of a spin-one system can be represented by eight linear equations in classical variables should not be surprising: the
  von Neumann equation is also of this form, where the eight variables are the eight independent elements of the $3\times3$ density matrix.}. $X_1, X_2, X_3$ represent the three spin components, while the remaining five variables represent nonlinear spin terms. These are effectively hidden variables in our approach. The advantage of our approach becomes apparent when we start coupling spin-one systems together and start doing approximations.

 Note that the eight equations~(\ref{bloch_group}) are not completely independent: they satisfy constraints set by conservation of the Casimir operators:
  \begin{align}
    \label{eq:Cas}
    C_1=\sum_{\alpha} X_\alpha^2,\; C_2=\sum_{\alpha\beta\gamma} d_{\alpha\beta\gamma} X_\alpha X_\beta X_\gamma,
  \end{align}
  where $d_{\alpha\beta\gamma}$ are the symmetric structure constants of the $SU(3)$ group.  Interestingly $X_1^2+X_2^2+X_3^2$, representing the sum of squares of the classical spin components, is not conserved under $SU(3)$ dynamics. There is no paradox here: recall all operators including $\hat S_{x,y,z}^2$ are represented through linear combination of operators $\hat X_\alpha$ so e.g. $X_1^2$ does not have the simple physical interpretation of $S_x^2$ even though $X_1$ does represent
  $S_x$.

\begin{figure}
  \includegraphics[width=0.7\linewidth]{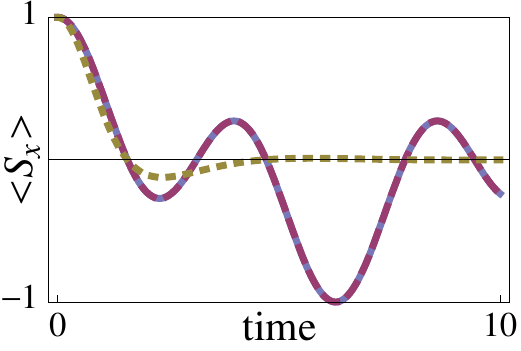}
  \caption{Comparison of the dynamics of $\langle \hat S_x \rangle$ for a spin-one particle initially pointed in the x-direction subjected to the Hamiltonian ${1\over 2} \hat S_z^2-\hat S_z$. The dynamics are calculated with exact diagonalization (solid, blue), $SU(3)$ TWA (dashed, red), and $SU(2)$ TWA (dotted, yellow). (Color online.)}
  \label{fig:onesite}
\end{figure}

To illustrate the difference between the ``naive'' $SU(2)$ TWA and the new $SU(3)$ TWA we consider the Hamiltonian~(\ref{one_spin_hamiltonian}) with $B_x=B_y=0$ and $B_z=1$. The $SU(2)$ and $SU(3)$ Weyl symbols corresponding to this Hamiltonian are \beq
(H_I)_W^{SU(2)}&=&(U/2) X_3^2-X_3-1/2,\nonumber\\
(H_I)_W^{SU(3)}&=&(U/6)(2-\sqrt{3} X_8)-X_3.  \eeq Note that in the $SU(2)$ case we chose the spin-one representation of the spin operators given by the first three operators of the $SU(3)$ representation. The additional constant term $-1/2$ in the $SU(2)$ Hamiltonian comes from $(\hat X_k^2)_W = X_k^2 - \tr(T_k^2)/4$. For concreteness we choose $U=1$ and start with the spin pointing along the $x$-direction and observe the expectation value of $\hat S_x$ as a function of time. In
Fig. \ref{fig:onesite} we show comparison of the resulting exact dynamics with $SU(2)$ and $SU(3)$ TWA approximations. As expected the $SU(3)$ TWA is exact while the $SU(2)$ semiclassical dynamics are only accurate at short times. The difference comes from the fact that any interaction terms in the $SU(2)$ case are represented by non-linearity while in the $SU(3)$ case they are represented by additional (hidden) variables, which in turn have their own quite complex dynamics.

Next let us consider a more complicated setup, where we deal with a system of interacting spin-one degrees of freedom such that the Hamiltonian becomes
\begin{align}
  \label{eq:spin_ham}
  \hat H = \sum_n \hat H_I^{(n)}+\hat H_C
\end{align}
where $H_I^{(n)}$ is the local spin-one Hamiltonian~(\ref{one_spin_hamiltonian}) describing $n$-th spin and
\begin{align}
  \hat H_C=- J \sum_{n\ne m}(\hat S_x^n \hat S_x^m +\hat S_y^n \hat S_y^m).
\end{align}
We have chosen a fully connected Hamiltonian to allow for comparison of TWA and exact dynamics for larger system sizes.

The Weyl symbol of the coupling term is the same for the $SU(2)$ and the $SU(3)$ representations because it does not involve local nonlinear spin-operators,
\begin{align}
  (H_C)_W = -J \sum_{n\ne m} (X_1^n X_1^m +X_2^n X_2^m).
\end{align}

For multi-spin systems we do not use the exact Wigner function, which is defined and integrated over the coherent state variables. Instead, we use a multivariate Gaussian distribution $f^n(X^n_1,\dots,X^n_{N^2-1})$ for each site $n$ and integrate over the $N^2-1$ $SU(N)$ variables:
\begin{align}
  \langle \hat \Omega(t)\rangle\approx\int \prod_n d X^n f^n(\vec X^n) \Omega_W(\vec X^{cl\,n}(t)),
\end{align}
where the mean and covariance matrix for each $f$ is fixed by the quantum expectation values of the initial state of the system, $\langle X^n_\alpha \rangle_{f^n} = \braket{\hat X^n_\alpha}$ and $\langle X^n_\alpha X^n_\beta \rangle_{f^n} = \braket{(\hat X^n_\alpha \hat X^n_\beta+\hat X^n_\beta \hat X^n_\alpha)/2}$ (see Supplementary Material for details). We use this best Gaussian approximation for two reasons. First, the exact Wigner function will in general have negative
values, so the integration depends on the cancellation of positive and negative contributions, which numerically requires more sample points to converge. Secondly, we numerically found that the best Gaussian TWA results are consistently more accurate. Formally this Gaussian scheme is justified if we increase the spin size (proportional to the conserved value of the Casimir operator). For an initial correlated (not product) state one should use the multivariate Gaussian which correctly reproduces
both local and non-local correlation functions like $\langle X^m_\alpha X^n_\beta \rangle$.  For observables, instead of Weyl ordering one can use direct quantum classical substitution $\hat X^n_m\to X^n_m$ because any onsite observable is linear in $\hat X$ and for linear operators this substitution is exact~\cite{Polkovnikov2010a}.

\begin{figure}
  \centering
  \includegraphics[width=0.45\linewidth]{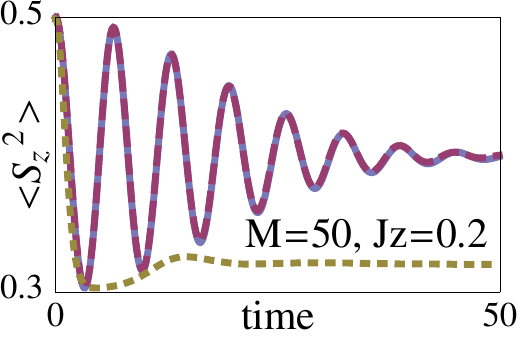}
  \includegraphics[width=0.45\linewidth]{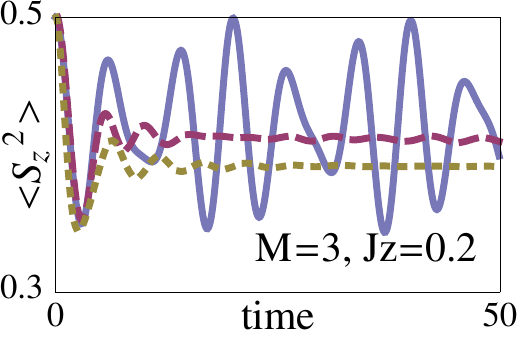}

  \includegraphics[width=0.45\linewidth]{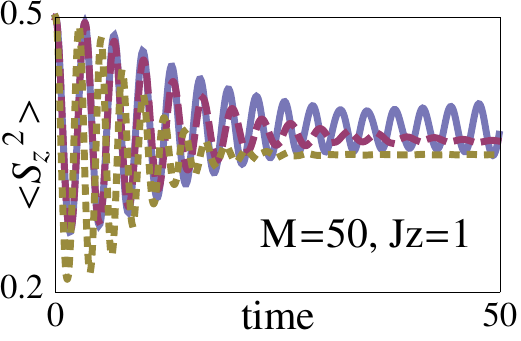}
  \includegraphics[width=0.45\linewidth]{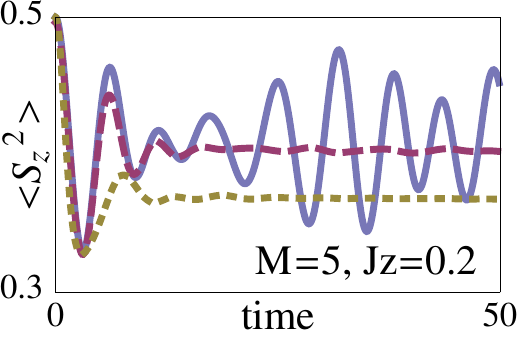}

  \includegraphics[width=0.45\linewidth]{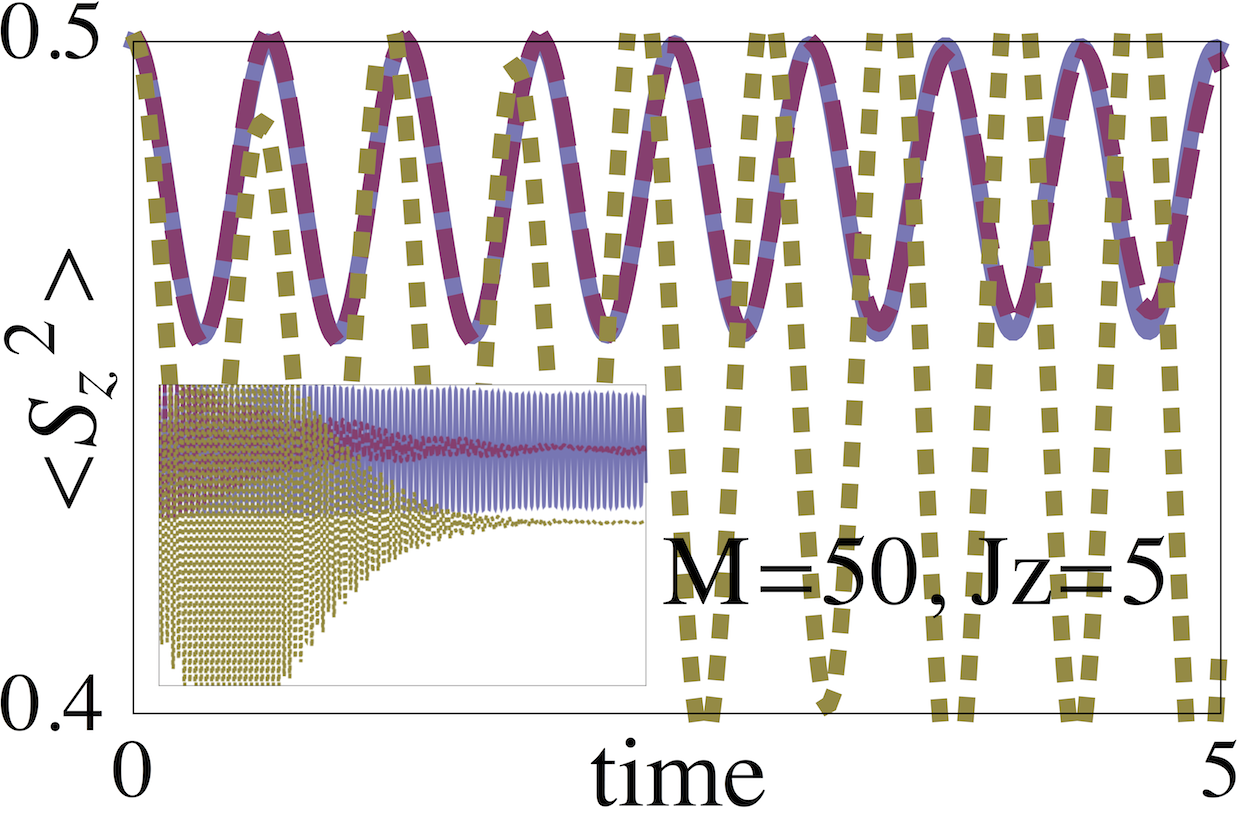}
  \includegraphics[width=0.45\linewidth]{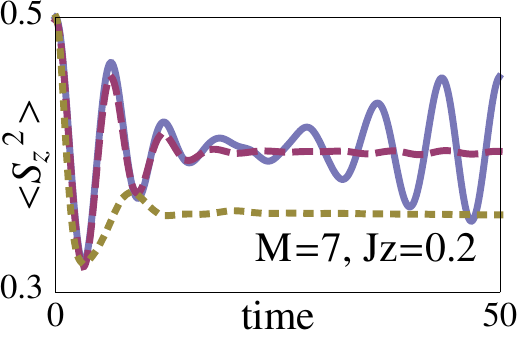}

  \caption{Comparison of the dynamics of $\langle \hat S^2_z \rangle$ for a fully connected system with $M$ spins due to Hamiltonian \eqref{eq:spin_ham} calculated with exact diagonalization (solid, blue), $SU(3)$ TWA (dashed, red), and $SU(2)$ TWA (dotted, yellow). Initially all spins are pointed in the $x$-direction.  On the left are the results for various coupling strengths $J z$, where $z=M-1$ is the connectedness (the inset in the bottom plot shows the same plot for times 0 to 50). On the
    right are the results for different system sizes. (Color online.)}
  \label{fig:fullycon}
\end{figure}

In Fig. \ref{fig:fullycon} we show the dynamics of the spin fluctuations $\langle \hat S_z^2 \rangle$ per site obtained by exact diagonalization, and $SU(2)$ and $SU(3)$ TWA. The system is initially prepared with all spins pointing in the $x$-direction.
We compare the dynamics for a fully connected system for different values of the coupling $J$ and for different system sizes. As the coupling is lowered and the on-site term in the Hamiltonian becomes more dominant, the $SU(3)$ TWA becomes a better approximation, while the $SU(2)$ becomes worse. When the on-site term is 5 times as dominant as the coupling term, the $SU(3)$ TWA is indistinguishable from exact quantum dynamics. As the system size increases, and hence each site is connected to more
sites, the $SU(3)$ TWA dynamics approach exact quantum dynamics. Similarly to the $SU(2)$ case, $SU(3)$ TWA fails to describe quantum revivals, which occur later and later in time as the system size increases.

\begin{figure}
  \centering
  \includegraphics[width=0.45\linewidth]{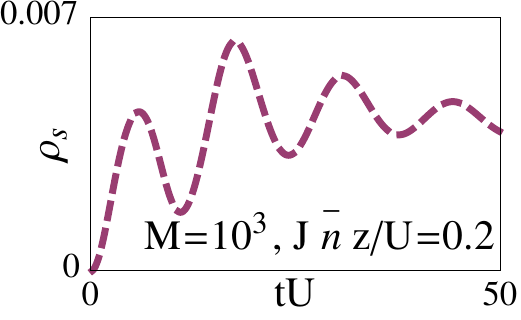}
  \includegraphics[width=0.45\linewidth]{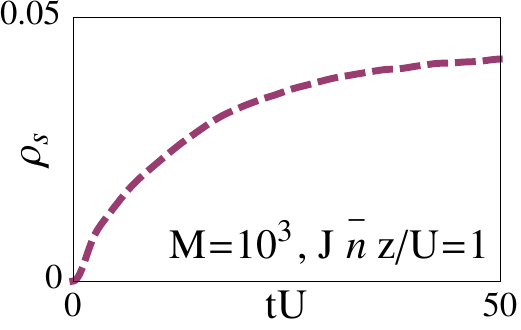}

  \quad \includegraphics[width=0.45\linewidth]{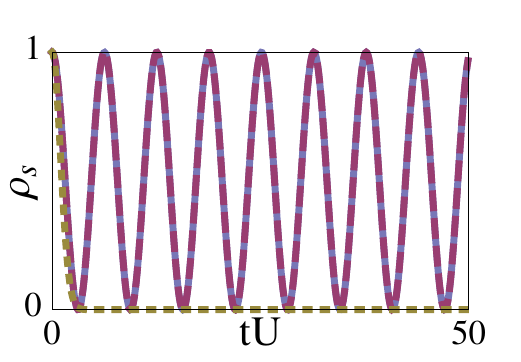}
  \includegraphics[width=0.45\linewidth]{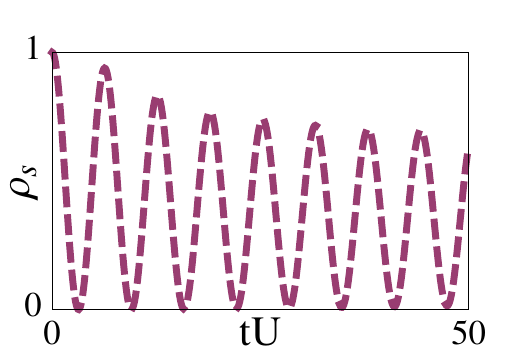}
  \caption{The dynamics of the order parameter $\rho_s$ for the effective Bose-Hubbard model \eqref{eq:BH_eff_ham}, calculated with $SU(3)$ TWA. The top row shows the results starting in the Mott insulator phase. The bottom row begins in the superfluid phase: in the left-hand plot, the system is quenched to $J=0$, while in the right hand-hand plot the system is quenched to $J \bar n z/U=0.1$. (Color online.)}
  \label{fig:quench}
\end{figure}

As a more practical example, we model the Bose-Hubbard model using the effective Hamiltonian~\cite{Altman2002}
\begin{align}
  \label{eq:BH_eff_ham}
  H_{eff} =& \frac{U}{2} \sum_i (\hat S_z^i)^2-J \bar n \sum_{\langle i j \rangle}(\hat S_x^i \hat S_x^j+\hat S_y^i \hat S_y^j)-\mu \sum_i \hat S_z^i,
\end{align}
where $\bar n$ is the mean particle density. This truncation of the Hilbert space to three dimensions per site is acceptable in the vicinity of the Mott insulating state~\cite{Huber2007}. We use $SU(3)$ TWA to determine the dynamics of the order parameter $\rho_s=\sum_{i \ne j} \langle \hat S^+_i \hat S^-_j\rangle/M^2$ (representing superfluid density) for a 3D system with $M=10^3$ sites in a cubic lattice with periodic boundary conditions. The results are plotted in Fig. \ref{fig:quench}.

First we quench from the Mott insulator phase, i.e. a Fock state on each site (the ground state for $J=0$). In terms of the effective spin Hamiltonian, this corresponds to a product state of $\ket{\hat S_z=0}$. The dynamics arise from an instantaneous quench to a finite coupling, either $J \bar n z/U=0.2$ or $J \bar n z/U=1$. In each case, the system moves away from a pure Mott insulator state; for a smaller coupling, there is some oscillation which is absent for a larger coupling.  The
superfluid density remains small, as a sudden quench leads to a high temperature state which does not exhibit long range order \cite{Polkovnikov2002a}.

We also show a quench from the superfluid phase (the ground state for $U=0$), which in terms of the effective spin Hamiltonian corresponds to a product state of $\ket{\hat S_x=+1}$. When the system is quenched to $J=0$, each site precesses independently. Thus we can calculate the dynamics using exact diagonalization, $SU(3)$ TWA, and $SU(2)$ TWA. Since the on-site Hamiltonian can be linearized in terms of $SU(3)$ variables, the $SU(3)$ TWA reproduces the exact quantum dynamics, including quantum
recurrences, while the $SU(2)$ TWA decays. When we instantaneously quench to $J \bar n z/U=0.1$, the $SU(3)$ TWA still reproduces the oscillations of quantum recurrences, damped by the coupling to the larger system.

In summary, we have introduced a semiclassical formalism for simulating the quantum dynamics of strongly interacting coupled-spin systems. We have shown that by increasing the phase space and introducing new (hidden) degrees of freedom one can partially account for local quantum fluctuations and significantly improve the accuracy of the semiclassical description of the dynamics. We have argued and shown numerically that the accuracy of this method increases as we increase connectivity of the
system. We have demonstrated numerically that this method accurately reproduced results of quench dynamics of coupled spin-one systems in a broad range of parameters including the strong coupling regime. As another illustration we analyzed quench dynamics across the superfluid-insulator transition in a three-dimensional Bose-Hubbard model.

While here we only presented results for $SU(3)$ variables, this formalism can be straightforwardly applied to any $SU(N)$ group (albeit with a larger phase space), where $N$ is the dimension of a local Hamiltonian. Thus we can use classical dynamics to exactly do local quantum dynamics which are linear in any $SU(N)$ representation. This should allow one to take into account quantum fluctuations within larger clusters and then use the TWA approximation to treat inter-cluster coupling. We are
planning to analyze this possibility in a future work. An important and open question is finding the optimal way of introducing hidden variables keeping their number larger than in the naive classical limit, yet much smaller than the Hilbert space size.

\acknowledgements We thank V. Oganesyan for the initial idea of increasing phase space and A.M. Rey for sharing details of unpublished work. This work was supported by AFOSR FA9550-313 13-1-0039, NSF DMR-1206410, and the Natural Sciences and Engineering Research Council of Canada.

\bibliography{library}

\clearpage

\begin{widetext}

\begin{center}
  \large{\bf Supplementary Material}
\end{center}

\end{widetext}

\section{Truncated Wigner Approximation}
\label{sec:phase-space-methods}

Perhaps the best known phase space representation is based on the
Wigner-Weyl quantization~\cite{Hillery1984, Polkovnikov2010a}, where
the Weyl symbol of an operator is represented through the partial
Fourier transform. In the coherent state representation corresponding
to second-quantized language one defines \be
\label{eq:11}
\Omega_W(\vec \alpha) = \int {d \vec \eta d\vec \eta^\ast\over 2}
\langle \vec \alpha - \vec \eta/2 | \hat \Omega | \vec \alpha + \vec
\eta/2 \rangle \mathrm e^{(\vec\eta^* \vec\alpha - \vec \eta \vec
  \alpha^*)/2}. \ee The Wigner function is simply the Weyl transform
of the density matrix. There are other equivalent representations of
the operators~\cite{gardiner-zoller, walls-milburn}, for example,
P-representation or Q-representation (also known as the Husimi
representation). For our purposes it is not essential which of these
representations is used but we will refer to Wigner-Weyl
quantization to be specific. Within phase space methods one can
describe expectation values of arbitrary operators as a standard
average, where the Wigner function $W(\alpha,\alpha^\ast)$ plays the
role analogous to the classical probability distribution: \be \langle
\Omega\rangle=\int d\vec\alpha d\vec\alpha^\ast
W(\vec\alpha,\vec\alpha^\ast) \Omega_W(\vec\alpha,\vec\alpha^\ast).
\ee

The equation of motion for the Wigner function
reads~\cite{Polkovnikov2010a} \be i\dot W =2 W\sin\left(\Lambda_c\over
  2\right) H_W, \ee where 
\begin{align}
  \label{sym}
  \Lambda_c=\sum_j
\frac{\overleftarrow \partial}{\partial \alpha_j} \frac{\overrightarrow\partial}{\partial \alpha_j^\ast}-
\frac{\overleftarrow \partial}{\partial \alpha_j^\ast}
\frac{\overrightarrow \partial}{\partial \alpha_j}
\end{align}
is the
so-called symplectic operator and $H_W$ is the Weyl symbol of the
Hamiltonian. In general this partial-differential equation is hard to
solve. However, it becomes simple if one can expand the $\sin$
function to leading order in $\Lambda_c$: \be i\dot W\approx W
\Lambda_c H_W.  \ee In this limit the right-hand side becomes just the
Poisson bracket of the Wigner function and the Hamiltonian. Therefore
the resulting equation is identical to the classical Liouville
equation for the probability distribution. We note that the factor of
$i$ comes from using the coherent state Poisson brackets (see
Ref.~\cite{Polkovnikov2010a} for details). In turn this Liouville
equation locally conserves the probability along the classical
trajectories described by the corresponding equations of motion 
\begin{align}
  \label{conclass}
  i\dot \alpha_{cl\,j}=\frac{\partial H_W}{\partial \alpha_{cl\,j}^\ast},
\end{align}
in this case the
Gross-Pitaevskii equations. The Wigner function simply describes the
distribution of initial conditions. This linearization of the
equations of motion (known as the truncated Wigner approximation
(TWA)~\cite{Steel1998}) can be justified near the classical limit,
where $\Lambda_c$ is small. For coherent states it is proportional to
$1/N$, where $N$ is the occupation of modes; in the
coordinate-momentum representation $\Lambda$ is explicitly
proportional to $\hbar$; and, for spin systems $\Lambda$ is inversely
proportional to the spin $S$. TWA also becomes exact for harmonic
systems, where the Hamiltonian $H_W$ does not contain terms higher
than quadratic in phase space variables. Indeed, in this case all terms
containing $\Lambda_c^3$ and higher will vanish because they contain
at least three derivatives.

\section{$SU(N)$ TWA}
\label{sec:sun-twa}

For a more detailed account of using specifically $SU(2)$ Schwinger
bosons in TWA,
see~\cite{Polkovnikov2010a} Section 3.3.

We can write operators with $SU(N)$ algebra using $N$ boson operator
pairs and the $N$ dimension matrix representation of the operators
$T^{ij}_\alpha$:
\begin{align}
  \label{genrep}
  \hat X_\alpha=\hat a^\dagger_i T_\alpha^{ij} \hat a_j.
\end{align}
We define the classical $SU(N)$ variables as the Wigner-Weyl transform
of the $SU(N)$ operators (recall the matrices are traceless):
\begin{align}
  \label{classgenrep}
  X_\alpha= \alpha^*_i T_\alpha^{ij} \alpha_j.
\end{align}
Using this as a definition, we can rewrite the symplectic operator
\eqref{sym} in
terms of the classical $SU(N)$ variables:
\begin{align}
  \label{eq:3}
  \Lambda_c = \frac{\overleftarrow \partial}{\partial X_\alpha} i
  f_{\alpha \beta \gamma} X_\gamma
  \frac{\overrightarrow \partial}{\partial X_\beta}.
\end{align}
In TWA, the dynamics are approximated to follow classical
trajectories, so the classical $SU(N)$ variables, as functions of the
classical canonical variables, simply follow the path determined by
Hamilton's equation:
\begin{align}
  \label{eq:6}
  \frac{\partial}{\partial t} X^{cl}_\alpha &= i H_W \Lambda_c X^{cl}_\alpha\\
  &= f_{\alpha \beta \gamma} \frac{\partial H_W}{\partial
    X^{cl}_\beta}X^{cl}_\gamma ,
\end{align}
where we have also rewritten $H_W$ in terms of classical $SU(N)$
variables.  We can see how this follows directly from the classical
equations of motion for the canonical variables; if we multiply
equation \eqref{conclass} by $\alpha^*_i T_\alpha^{ij}$, and the
complex conjugate equation by $T_\alpha^{ji}\alpha_i$ (where the $\alpha$ represent classical paths), their
addition yields the time dependency of the $SU(N)$ variables:
\begin{align}
  \label{eq:16}
  \frac{\partial}{\partial t} X^{cl}_\alpha &= \alpha^*_i
  T_\alpha^{ij}\frac{\partial \alpha_j}{\partial t} + \frac{\partial
    \alpha_i^*}{\partial t}
  T_\alpha^{ij}\alpha_j \\
  &= -i \alpha^*_i T_\alpha^{ij}\frac{\partial
    H_W}{\partial \alpha_j^\ast} + i
  \frac{\partial H_W}{\partial \alpha_i}  T_\alpha^{ij}\alpha_j.
\end{align}
Taking the derivative of equation \eqref{classgenrep} to get
\begin{align}
  \label{eq:17}
  \frac{\partial}{\partial \alpha_j} = \alpha^*_i T_\alpha^{ij}\frac{\partial}{\partial X^{cl}_\alpha},
\end{align}
we get
\begin{align}
  \label{eq:18}
  \frac{\partial}{\partial t} X^{cl}_\alpha &=-i \alpha^*_i  T_\alpha^{ij}T_\beta^{jk}\alpha_k\frac{\partial
    H_W}{\partial X_\beta} + i
  \frac{\partial H_W}{\partial  X_\beta}\alpha^*_k T_\beta^{ki}
  T_\alpha^{ij}\alpha_j\\
  &=-i \alpha^*_i\left[T_\alpha,T_\beta\right]^{ij}\alpha_j
  \frac{\partial H_W}{\partial  X^{cl}_\beta} \\
  &= f_{\alpha \beta \gamma} X^{cl}_\gamma  \frac{\partial H_W}{\partial  X^{cl}_\beta}.
\end{align}

\section{$SU(3)$ Schwinger Bosons }
\label{sec:su3-schwinger-bosons}

We use a fundamental representation of $SU(3)$ with the following
basis matrices (similar to \cite{Kiselev2013}):

\begin{gather}
  T_1 = \left(
    \begin{array}{ccc}
      0 & \frac{1}{\sqrt{2}} & 0 \\
      \frac{1}{\sqrt{2}} & 0 & \frac{1}{\sqrt{2}} \\
      0 & \frac{1}{\sqrt{2}} & 0 \\
    \end{array}
  \right),T_2=\left(
    \begin{array}{ccc}
      0 & -\frac{i}{\sqrt{2}} & 0 \\
      \frac{i}{\sqrt{2}} & 0 & -\frac{i}{\sqrt{2}} \\
      0 & \frac{i}{\sqrt{2}} & 0 \\
    \end{array}
  \right),\\
  T_3=\left(
    \begin{array}{ccc}
      1 & 0 & 0 \\
      0 & 0 & 0 \\
      0 & 0 & -1 \\
    \end{array}
  \right),T_4=\left(
    \begin{array}{ccc}
      0 & 0 & 1 \\
      0 & 0 & 0 \\
      1 & 0 & 0 \\
    \end{array}
  \right),\\
  T_5=\left(
    \begin{array}{ccc}
      0 & 0 & -i \\
      0 & 0 & 0 \\
      i & 0 & 0 \\
    \end{array}
  \right),T_6=\left(
    \begin{array}{ccc}
      0 & -\frac{1}{\sqrt{2}} & 0 \\
      -\frac{1}{\sqrt{2}} & 0 & \frac{1}{\sqrt{2}} \\
      0 & \frac{1}{\sqrt{2}} & 0 \\
    \end{array}
  \right),\\
  T_7=\left(
    \begin{array}{ccc}
      0 & \frac{i}{\sqrt{2}} & 0 \\
      -\frac{i}{\sqrt{2}} & 0 & -\frac{i}{\sqrt{2}} \\
      0 & \frac{i}{\sqrt{2}} & 0 \\
    \end{array}
  \right),T_8=\left(
    \begin{array}{ccc}
      -\frac{1}{\sqrt{3}} & 0 & 0 \\
      0 & \frac{2}{\sqrt{3}} & 0 \\
      0 & 0 & -\frac{1}{\sqrt{3}} \\
    \end{array}
  \right).
\end{gather}
Note that these are normalized for spin-one, such that
\begin{align}
  \label{eq:2}
  \tr(T_iT_j)=2\delta_{ij}.
\end{align}

The non-zero structure constants are
\begin{gather}
  \label{eq:5}
  f_{123}=f_{147}=f_{165}=f_{246}=f_{257}=f_{367}=1,\\
  f_{178}=f_{286}=\sqrt{3},\\
  f_{345}=2,
\end{gather}
where the rest can be determined by permuting the indices (they are
ant-symmetric under permutations).

The non-zero symmetric structure constants are
\begin{gather}
  \label{eq:5}
  d_{114}=d_{125}=d_{477}=1,\\
  d_{136}=d_{224}=d_{237}=d_{466}=d_{567}=-1,\\
  d_{118}=d_{228}=d_{668}=d_{778}=\frac{1}{\sqrt{3}},\\
  d_{338}=d_{448}=d_{558}=d_{888}=-\frac{2}{\sqrt{3}},
\end{gather}
where the rest can be determined by permuting the indices (they are
symmetric under permutations).

Since the first three matrices have the algebra of $SU(2)$, they can
be used to construct Schwinger bosons that can represent any magnitude
of spin. We have the mapping
\begin{gather}
  T_1 =  \hat S_x\\
  T_2 = \hat S_y\\
  T_3 = \hat S_z.
\end{gather}

For the special case of spin-one, we can construct any operator with all
the appropriate algebra as a linear combination of the eight $SU(3)$
matrices. The generators themselves are related to spin-one operators as
follows:
\begin{gather}
  T_4 =  (S_x)^2- (S_y)^2\\
  T_5 = [S_x,S_y]_+\\
  T_6 = [S_x,S_z]_+\\
  T_7 = [S_y,S_z]_+\\
  T_8 = \frac{1}{\sqrt{3}}\left((S_x)^2+ (S_y)^2-2 (S_z)^2\right),
\end{gather}
where the hats have been left off the spins to emphasize that these
equalities are only true for the spin-one spin matrices, not the spin
operators in general.

For example, we can represent the squares of the spins as
\begin{gather}
  (S_x)^2 = \frac{1}{3}(2 \mathbb{1} + \frac{\sqrt{3}}{2} T_8)+\frac{1}{2}T_4\\
  (S_y)^2 = \frac{1}{3}(2 \mathbb{1} + \frac{\sqrt{3}}{2} T_8)-\frac{1}{2}T_4\\
  (S_z)^2 = \frac{1}{3}(2 \mathbb{1} - \sqrt{3} T_8).
\end{gather}

\section{Gaussian approximate Wigner functions}
\label{sec:gauss-appr-wign}

Using Gaussian distributions instead of the exact Wigner functions does
not necessarily worsen the approximation, since the difference between
the exact Wigner function and the appropriately chosen Gaussian
distribution approximation is of the same order in $1/S$ as the
corrections to the TWA dynamics. Formally one can understand this by
requiring that expectation values of the spin variables $\hat
X_1,\dots \hat X_8$ and their variances agree with those given by the
Gaussian up to order $1/S^2$, where $S$ is the spin size
(proportional to the conserved Casimir operator as we discuss
next). The Gaussian distribution has a clear advantage in that it is
always positive and thus easy to handle. Numerically we found that the
accuracy of TWA with the Gaussian distribution is consistently better
than that of the TWA with the exact Wigner function.

Note that the initial states we choose are all product states, so the Wigner
functions are simply a product of single site Wigner
functions. We do this for simplicity; the initial Wigner function can
also represent an entangled initial state.

In the Gaussian approximation, then, the expectation value for an
operator $\hat \Omega$ will be 
\begin{align}
  \label{eq:10}
  \braket{\hat \Omega} \approx \int \prod_m\prod_k dX_k^m \frac{e^{-(R_{kl}
      X_l^m-\mu_k)^2/2 \sigma_k^2}}{\sqrt{2 \pi} \sigma_k} \Omega_W,
\end{align}
where $m$ runs over the number of sites and $k$ over the eight $SU(3)$ variables.

The expectation values of the means and symmetrized correlation
functions, which we require to exactly match those in the initial
state, are given on each site by
\begin{gather}
  \label{eq:4}
  \tr(\hat \rho_0 \hat T_i) = \int \prod_k dX_k \frac{e^{-(R_{kl}
      X_l-\mu_k)^2/2 \sigma_k^2}}{\sqrt{2 \pi} \sigma_k} X_i
\end{gather}
and
\begin{multline}
  \label{eq:4}
  \tr(\hat \rho_0 \frac{1}{2}(\hat T_i \hat T_j + \hat T_j \hat T_i)) =\\
  \int \prod_k dX_k \frac{e^{-(R_{kl} X_l-\mu_k)^2/2
      \sigma_k^2}}{\sqrt{2 \pi} \sigma_k} X_i X_j,
\end{multline}
where $R,\mu$ and $\sigma$ are free parameters of the Gaussian, which
we determine from the initial conditions.

For example, for $\rho_0=\ket{\hat S_x=1}\bra{\hat S_x=1}$, we rotate
with
\begin{align}
  \label{eq:8}
  R=
  \begin{pmatrix}
    0 & 0 & 0 & -\frac{1}{2} & 0 & 0 & 0 & \frac{\sqrt{3}}{2} \\
    0 & 0 & 0 & 0 & 0 & 0 & 1 & 0 \\
    0 & 0 & -\frac{1}{\sqrt{2}} & 0 & 0 & \frac{1}{\sqrt{2}} & 0 & 0 \\
    0 & \frac{1}{\sqrt{2}} & 0 & 0 & \frac{1}{\sqrt{2}} & 0 & 0 & 0 \\
    0 & 0 & 0 & \frac{\sqrt{3}}{2} & 0 & 0 & 0 & \frac{1}{2} \\
    0 & 0 & \frac{1}{\sqrt{2}} & 0 & 0 & \frac{1}{\sqrt{2}} & 0 & 0 \\
    0 & -\frac{1}{\sqrt{2}} & 0 & 0 & \frac{1}{\sqrt{2}} & 0 & 0 & 0 \\
    1 & 0 & 0 & 0 & 0 & 0 & 0 & 0 \\
  \end{pmatrix}
\end{align}
and have Gaussians determined by
\begin{gather}
  \label{eq:9}
  \mu =
  \begin{pmatrix}
    0\\
    0\\
    0\\
    0\\
    \frac{1}{\sqrt{3}}\\
    0\\
    0\\
    1
  \end{pmatrix}
  \quad \sigma=
  \begin{pmatrix}
    1\\
    1\\
    1\\
    1\\
    0\\
    0\\
    0\\
    0
  \end{pmatrix}.
\end{gather}

Note that alternatively one can write the multivariate Gaussian
distribution in a more conventional form
\begin{align}
  \label{eq:7}
  f(X_1,\dots,X_8)=\frac{1}{\sqrt{(2
      \pi)^8|\Sigma|}}e^{-(X_i-\mu_i)\Sigma^{-1}_{ij}(X_j-\mu_j)}.
\end{align}
Here the vector $\mu_j$ is the vector of means and $\Sigma$ is the
covariance matrix.

\section{Casimir operators}
\label{sec:casimir-operators}

The conservation of the classical form of the two Casimir operators is
made plain when they are written in terms of the conjugate variables:
\begin{gather}
  \label{eq:1}
  C_1=\sum_{\alpha=1}^b X_\alpha^2=\frac{4}{3}\left(|\alpha_1|^2+|\alpha_2|^2+|\alpha_3|^2\right)^2, \\
  C_2=\sum_{\alpha\beta\gamma} d_{\alpha\beta\gamma} X_\alpha X_\beta
  X_\gamma=\frac{16}{9}\left(|\alpha_1|^2+|\alpha_2|^2+|\alpha_3|^2\right)^3.
\end{gather}
The conservation of the Casimir operators is assured by conservation
of the total number of Schwinger bosons; conservation of total number
of Schwinger bosons is guaranteed by the global $U(1)$ symmetry of the
Hamiltonian $\alpha_i \rightarrow \alpha_i e^{i \theta}$ which is
imposed by only including conjugate variables as they are combined in
$SU(3)$ operators. Note that the Casimir operators become larger as we
approach the classical limit so the inverse of say $C_1$ determines
the strength of quantum fluctuations in the system.

\end{document}